\documentclass[osajnl,twocolumn,showpacs,superscriptaddress,10pt]{revtex4-1} 
\usepackage{amsmath,amssymb,graphicx}
\begin{document}

\title{Kerr-lens Mode Locking Without Nonlinear Astigmatism}

\author{Shai Yefet}
\author{Valery Jouravsky}
\author{Avi Pe'er}\email{Corresponding author: Avi.Peer@biu.ac.il}
\affiliation{Department of physics and BINA Center of nano-technology, Bar-Ilan university, Ramat-Gan 52900, Israel}

\begin{abstract}We demonstrate a Kerr-lens mode locked folded cavity using a planar (non-Brewster) Ti:sapphire crystal as a gain and Kerr medium, thus cancelling the nonlinear astigmatism caused by a Brewster cut Kerr medium. Our method uses a novel cavity folding in which the intra-cavity laser beam propagates in two perpendicular planes such that the astigmatism of one mirror is compensated by the other mirror, enabling the introduction of an astigmatic free, planar-cut gain medium. We demonstrate that this configuration is inherently free of nonlinear astigmatism, which in standard cavity folding needs a special power specific compensation.
\end{abstract}


\maketitle 

\section{Introduction}

Astigmatism is a well known aberration in folded optical cavities that include Brewster-cut crystals and/or off-axis focusing elements \cite{HechtEugeneOptics}. For continuous-wave (CW) operation, astigmatism is linear and can be fully compensated by correctly choosing the folding angles of the focusing elements in accordance with the length of the Brewster-cut windows \cite{astigmatismcompensation}. This results in a circular non-astigmatic beam at the output of the laser. For mode-locked (ML) operation induced by the nonlinear Kerr effect, an additional nonlinear astigmatism from the Kerr lens is added that is power-dependent and needs to be taken into account. The standard technique to compensate for nonlinear astigmatism, is to deliberately introduce linear astigmatism in the "opposite" direction \cite{stabilitydesign}, which results in compensation of the overall astigmatism for a specific intra-cavity intensity. Changing any of the cavity parameters that affect the intra-cavity intensity will require recompensation. Here we demonstrate a novel type of cavity folding that eliminates from the source nonlinear astigmatism in Kerr-lens ML lasers.

\section{Standard cavity design}

\begin{figure}
\centerline{\includegraphics[width=8cm
]
{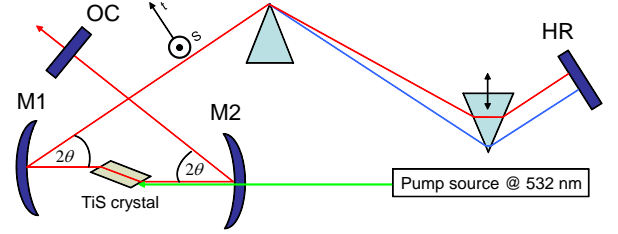}}
\caption{\label{standardfolding} Standard X-fold configuration of a TiS cavity. In this configuration, through the entire cavity the sagittal/tangential component of the laser mode is perpendicular/parallel to the optical table.}
\end{figure}

Let us first review shortly the standard design of a ML TiS cavity, illustrated in Fig.\ref{standardfolding}. The focusing mirrors are tilted with respect to the beam propagation axis, forming an X-fold (or Z-fold) cavity keeping the laser beam propagation parallel to the optical table. In this configuration, the astigmatism of the Brewster-cut crystal compensates for astigmatism of the curved mirrors.

The most important parameter for analyzing ML cavities is the strength of the Kerr effect defined as \cite{gammadefinition}:

\begin{equation}
\gamma=\frac{P_{c}}{\omega}\left(\frac{d\omega}{dP}\right)_{P=0} , \label{gamma}
\end{equation} where $P$ is the intra-cavity peak power, $\omega$ is the mode radius at the OC for a given distance between $M1$ and $M2$ and $P_{c}$ is the critical power for catastrophic self-focusing \cite{criticalpower}. From Eq.\ref{gamma} it follows that the Kerr lens strength is represented by the change of the mode size on the OC due to a small increase in the intra-cavity peak power. This dependence of the mode size is due to the self-focusing effect caused by the intensity dependent refractive index of the crystal: $n=n_{0}+n_{2}I=n_{0}+n_{2}P/A$, where $n_{0}$ and $n_{2}$ are the zero order and second order refractive indices respectively \cite{valueofn2}, and $A$ is the mode area.

\begin{figure}
\centerline{\includegraphics[width=8cm
]
{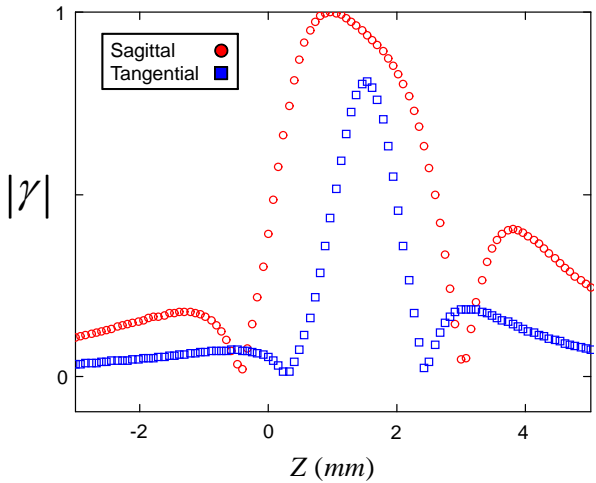}}
\caption{\label{gammaforstandardfolding} Normalized Kerr-lens strength $|\gamma|$ in X-fold configuration as a function of the crystal position $Z$ for the sagittal (red circles) and tangential (blue squares) planes. At $Z=0$ the center of the crystal is separated by $R/2$ from $M2$. This calculation was performed for the cavity of Fig.\ref{standardfolding} including a $L=5mm$ long Brewster-cut crystal bounded between two focusing mirrors with $R=15cm$. The short arm between $M2$ and the output coupler is $30cm$ and the long arm between $M1$ and the high reflector (HR) is $90cm$.}
\end{figure}

To determine the working point for ML we define $\delta$ as the distance between $M1$ and $M2$ with respect to an arbitrary reference point. Two separate bands of $\delta$ values $[\delta_{1},\delta_{2}],\ [\delta_{3},\delta_{4}]$ allow stable CW operation of the cavity \cite{KLM}. These two stability zones are bounded by four stability limits ($\delta_{4}>\delta_{3}>\delta_{2}>\delta_{1}$), each one requires different angle values to compensate for linear astigmatism in CW operation. Astigmatic cavities are usually analyzed by splitting the cavity into tangential and sagittal planes, where both stable CW solution and $\gamma$ are calculated for each plane separately. We note that while for CW operation the uncoupled resonators calculation is quantitatively accurate, it is only qualitatively relevant for mode-locked operation since the sagittal and tangential planes are coupled as the beam propagates through the Kerr medium \cite{astigmaticselffocusing} and the change of the beam size in one plane affects the lens strength also in the other plane. As we show hereon, this coupling is effectively nulled in the novel cavity folding, making the uncoupled resonators calculation quantitatively accurate.

It was shown \cite{KLMnearstabilitylimits} that $|\gamma|$ is maximized close to the cavity stability limits. We therefore calculated $|\gamma|$ using the transformed complex beam parameter method presented by \cite{inverseqtransform} near the second stability limit $\delta_{2}$ at a position corresponding to a beam size with a typical diameter of $2.45mm$. Calculations are performed within the aberration free Kerr lens approximation, in which the transverse variation of the refractive index is approximated to be parabolic, so that the beam maintains its Gaussian shape during the propagation and ABCD method to analyse the cavity can be applied.

Figure \ref{gammaforstandardfolding} plots the normalized $|\gamma|$ of the sagittal and tangential planes for the standard cavity folding as a function of the crystal position $Z$. The folding angles were chosen so that linear CW astigmatism is compensated resulting in a circular CW beam on the OC. As the beam refracts into the crystal at Brewster angle, the mode size in the tangential plane increases by a factor of $n_{0}=1.76$ whereas in the sagittal plane it does not change. Since $|\gamma|$ depends on the strength of self-focusing, which in turn depends on the intra-cavity intensity $I$, a different mode size in each plane will cause a difference in the self-focusing strength, thus resulting in an astigmatic Kerr lens. It is clear that a linearly compensated CW beam will not remain circular after ML.

For this reason, cavities that produce circular beams in ML operation require the initial CW beam to be deliberately astigmatic so that the plane with the stronger $|\gamma|$ will "catch up" with the weaker one \cite{stabilitydesign}. However, a solution that turns an astigmatic CW beam to a non-astigmatic ML beam exists only for a specific value of the intensity $I$. Any change in parameters that keeps the CW astigmatism compensated but affects the intra-cavity intensity (peak power or mode size in the crystal) will require a change in the folding angles to match the precise CW astigmatism needed to converge into a non-astigmatic ML beam. This includes a change in: pump power, pump focusing, output coupler, short arm length and also $Z$ or $\delta$. Another disadvantage of the standard folding is that for long crystals, only a small fraction of the crystal contributes to the ML process, deteriorating the overall pump-mode overlap and affecting the laser efficiency and robustness.


\section{Novel cavity design}

\begin{figure}
\centerline{\includegraphics[width=8cm
]
{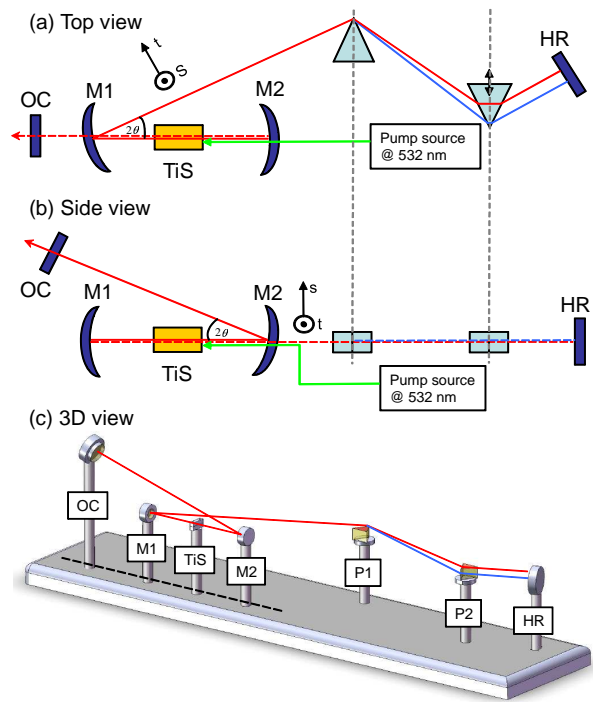}}
\caption{\label{novelconfiguration} Novel configuration of the TiS cavity in (a) top view, (b) side view and (c) 3D view of the optical table. The long arm remains parallel to the optical table, while the short arm is raised above the optical table. The illustrated sagittal and tangential components corresponds to the long arm with respect $M1$, while they switch in the short arm with respect to $M2$.}
\end{figure}

\begin{figure}
\centerline{\includegraphics[width=8cm
]
{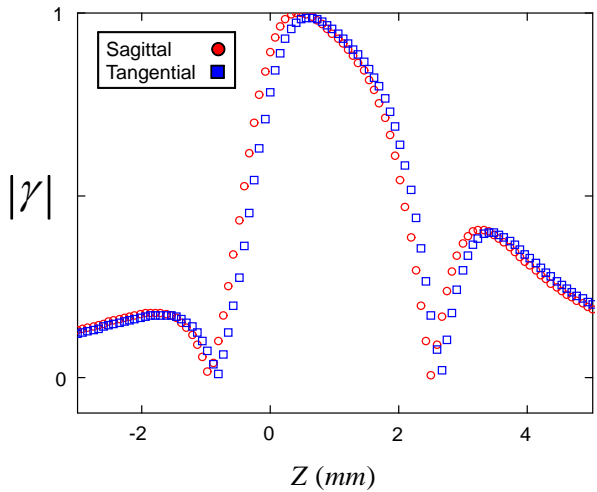}}
\caption{\label{gammafornovelfolding} Normalized Kerr-lens strength in novel configuration as a function of the crystal position for the sagittal (red circles) and tangential (blue squares) planes.}
\end{figure}

Here we demonstrate a different type of cavity folding which allows the introduction of a planar-cut (non-Brewster) crystal where the laser beam enters the crystal at normal incidence. In this configuration, the curved mirrors compensate for the astigmatism of each other instead of the crystal, and the spatial mode of the beam in both planes does not change as it enters the crystal, thus cancelling the nonlinear Kerr-lens astigmatism from the source. Fig.\ref{novelconfiguration} illustrates the laser configuration. Mirror $M1$ folds the beam in-plane of the optical table, whereas $M2$ folds the beam upwards. Thus, the sagittal and tangential components of $M1$ exchange roles at $M2$ leading to exact cancellation of the linear astigmatism of one mirror by the other mirror. Note that the polarization of the laser is unaffected by the new folding, which interchanges the primary axes, but does not mix them. The normalized Kerr-lens strength for this configuration is plotted in Fig.\ref{gammafornovelfolding} for the same $\delta$ as in the standard folding (other cavity parameters remain unchanged). It is clear that the Kerr-lens strength of each plane is equal with a small separation between the curves of each plane due to the non-zero value of the angles, taken to be $\theta\approx2.5^{o}$, the minimal value possible with the opto-mechanical mounts in our experiment.

We note that in this configuration, the mirrors astigmatism is uncoupled from the astigmatism of the Brewster-cut windows. In order to completely compensate for linear astigmatism, the slight astigmatism from the Brewster-cut prism-pair must be taken into account. Thus, equal folding angles will compensate for astigmatism only when the beam is collimated in both arms (at $\delta_{1}$). For $\delta_{2}$ however, equal angles can still be maintained using the prisms astigmatism to compensate for the residual mirrors astigmatism at $\delta_{2}$ by correctly choosing the overall propagation length inside the prisms.

Experimentally, it was critical to slightly tilt the crystal so that reflections from its surface will not interrupt the ML process. We measured that for ML to operate, the minimal incidence angle of the laser beam at the crystal surface was $\approx3.7^{o}$ which causes a slight increase in the tangential mode size inside the crystal, thus resulting in a negligibly  small nonlinear astigmatism. To estimate the nonlinear astigmatic effect due to the crystal tilting angle, we compare the decrease in the tangential optical path in the crystal with respect to that of the sagittal being $L/n_{0}$. Using ABCD matrices \cite{ABCDfortiltedinterface}, the decrease factor in the tangential plane for Brewster angle can be calculated to be $1/n_{0}^{2}\approx0.32$ , while for $3.7^{o}$, this factor becomes $0.997$ and can be completely neglected.

\section{Results}

\begin{figure}
\centerline{\includegraphics[width=8cm
]
{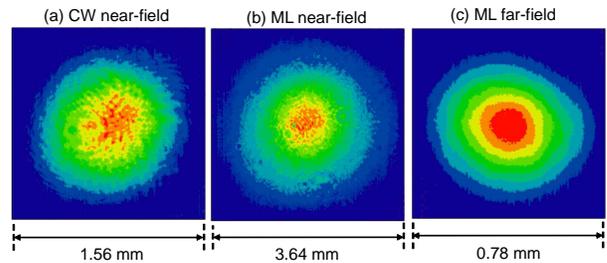}}
\caption{\label{spatialmodes} Intensity profiles of: (a) CW near-field, (b) ML near-field, (c) ML far-field. The near-field profile was taken at the OC and the far-field profile was taken at the focus of a $100cm$ positive lens.}
\end{figure}

The cavity was mode locked using pump power of $4.6W$ focused to a diameter of $45\mu m$ in the gain medium ($5mm$ long, $0.25$ wt$\%$ doped) with an OC of $85\%$. A prism pair of BK7 glass with $40cm$ separation was used for dispersion compensation. The cavity produced $\approx480mW$ of pulse output power with spectral bandwidth of $\approx100nm$. We note that the pulse bandwidth is not fully optimized and broader bandwidths can be achieved with better dispersion compensation.

Figure \ref{spatialmodes} plots the intensity profile of the CW and ML beams in the near field and for the ML beam also in the far field as measured by a CCD camera. For the near field, the CW astigmatism (defined as the ratio between the radius of the sagittal ($\omega_{s}$) and tangential ($\omega_{t}$) components) was fitted to be $0.98$, in good agreement to our prediction and the beam has an average radius of $0.5(\omega_{s}+\omega_{t})=0.66mm$. As expected, the ML astigmatism remains exactly the same with an average beam radius of $1.38mm$. The focusing quality of the ML mode was measured in the far field showing a beam quality factor of $M^{2}=1.6$.

We note that using a planar cut crystal reduces the mode size inside the crystal compared to a Brewster-cut crystal, since the beam does not expand in one dimension upon refraction into the crystal. This can reduce the laser threshold and increase the pulse output power compared to a standard cavity folding of similar parameters. In addition, the reduced mode size enhances the nonlinear Kerr lensing compared to Brewster-cut crystals leading to an increased mode locking strength. The large difference in beam size between the ML and CW modes (Fig.\ref{spatialmodes}) is a result of the improved Kerr strength, demonstrating the higher efficiency of the Kerr nonlinear mechanism in this design.

\section{Conclusion}

To conclude, we demonstrated a cavity design that in the first place needs no compensation for nonlinear astigmatism. In addition, the range of the crystal position in which both sagittal and tangential components contribute for the ML process is increased, resulting in a simple robust configuration for mode-locked lasers.

Special thanks to Mr. Shimon Pilo for illustrations and technical assistance. This research was supported by the Israeli science foundation (grant 807/09) and by the Kahn foundation.


\end{document}